\documentclass[manuscript,screen]{acmart}
\AtBeginDocument{%
  }
\PassOptionsToPackage{pdftex}{graphics}

\setcopyright{acmlicensed}
\copyrightyear{2018}

\usepackage{amsmath,amssymb,amsfonts}

\usepackage{booktabs}
\usepackage{multirow}
\usepackage{array}
\usepackage{makecell}
\usepackage{tabularx}

\usepackage{algorithmic}
\usepackage{listings}
\usepackage{xcolor}

\usepackage{float}
\usepackage{wrapfig}
\usepackage{picinpar}

\usepackage{pdfpages}
\usepackage{url}
\usepackage{hyperref}

\usepackage{soul,xcolor}
\usepackage{bbding}
\usepackage{pifont}
\usepackage{wasysym}
\usepackage{caption}

\usepackage{spotcolor}
\usepackage{xparse}
\usepackage{blindtext}
\usepackage{morewrites}
\usepackage{xkeyval}
\usepackage{natbib}

\definecolor{lightblue_qing}{RGB}{47,165,212}

\newcolumntype{L}[1]{>{\raggedright\arraybackslash}p{#1}}
\newcolumntype{C}[1]{>{\centering\arraybackslash}p{#1}}

\hypersetup{
    colorlinks=true,
    linkcolor=lightblue_qing,
    filecolor=magenta,      
    urlcolor=cyan,
    citecolor=lightblue_qing,
    pdftitle={Overleaf Example},
    pdfpagemode=FullScreen,
}

\acmYear{2018}
\acmDOI{XXXXXXX.XXXXXXX}
\acmConference[Conference acronym 'XX]{Make sure to enter the correct
  conference title from your rights confirmation email}{June 03--05,
  2018}{Woodstock, NY}
\acmISBN{978-1-4503-XXXX-X/2018/06}

\begin{document}

\title{MoveScanner: Analysis of Security Risks of Move Smart Contracts}

\author{Yuhe Luo}
\affiliation{
  \institution{Hainan University}
  \country{China}
}
\email{2213312852@qq.com}
\authornote{Both Yuhe Luo and Zhongwen Li are co-first authors.}
\author{Zhongwen Li}
\authornotemark[1]
\affiliation{%
  \institution{Hainan University}
  \country{China}
}
 \email{lzw123@hainanu.edu.cn}

\author{Xiaoqi Li}
\affiliation{
  \institution{ Hainan University}
  \country{China}
}
\email{csxqli@ieee.org}


\begin{abstract}

  As blockchain technology continues to evolve, the security of smart contracts has increasingly drawn attention from both academia and industry. The Move language, with its unique resource model and linear type system, provides a solid foundation for the security of digital assets. However, smart contracts still face new security challenges due to developer programming errors and the potential risks associated with cross-module interactions. This paper systematically analyzes the limitations of existing security tools within the Move ecosystem and reveals their unique vulnerability patterns. To address these issues, it introduces MoveScanner, a static analysis tool based on a control flow graph and data flow analysis architecture. By incorporating cross-module call graph tracking, MoveScanner can effectively identify five key types of security vulnerabilities, including resource leaks, weak permission management, and arithmetic overflows.
In terms of design, MoveScanner adheres to a modular principle, supports bytecode-level analysis and multi-chain adaptation, and introduces innovative resource trajectory tracking algorithms and capability matrix analysis methods, thereby significantly reducing the false positive rate. Empirical results show that MoveScanner achieved 88.2\% detection accuracy in benchmark testing, filling the gap in security tools in the Move ecosystem. Furthermore, this paper identifies twelve new types of security risks based on the resource-oriented programming paradigm and provides a theoretical foundation and practical experience for the development of smart contract security mechanisms. Future work will focus on combining formal verification and dynamic analysis techniques to build a security protection framework covering the entire contract lifecycle.
\end{abstract}


\maketitle

\section{Introduction}
\

Blockchain technology has evolved continuously since the birth of Bitcoin, from the introduction of smart contracts
by Ethereum, and the emergence of third-generation blockchains such as Solana, Aptos, and Sui in recent years. It has
gradually transitioned from a single-purpose value-storage system to a programmable distributed application platform.
Smart contracts, as a core component of blockchain, grant the system programmability but also introduce new security
risks\cite{luu2016making}. The Move language, proposed by the Meta team, was designed with security as its core objective. Its innovative resource model treats digital assets as first-class citizens, achieving non-replicability and non-discardability through a
linear type system and explicit move operations, thereby preventing resource leaks and double spending at the language
level. Combined with a strict type system, formal verification\cite{gong2025information}, and fine-grained permission controls, Move offers
superior security compared to traditional contract languages, such as Ethereum. However, in actual development, Move
smart contracts may still contain vulnerabilities due to programming errors, complex business logic, or cross-module
interactions\cite{tsankov2018securify}.

According to Certik’s 2023 Blockchain Security Report, losses caused by smart contract vulnerabilities exceeded
\$2.83 billion, with re-entrancy attacks, permission vulnerabilities, and logical errors as the primary attack types\cite{liu2024overview}.
The Ethereum ecosystem has developed a relatively mature security analysis toolchain, including Oyente, Securify,
ZEUS, Slither, and Mythril, which utilize symbolic execution, abstract interpretation, or static analysis methods to this study lay the foundation for smart contract security research\cite{li2025scalm}. In contrast, security tools for the Move ecosystem remain underdeveloped. Move Prover is primarily focused on the formal verification of specific properties, while MoveCheck and
VerMove can detect some errors but struggles with complex cross-module vulnerabilities. Manual audit services provided
by security companies also lack automation and systematic approaches\cite{tolmach2021survey}. This highlights the lack of comprehensive,
automated security analysis tools for smart contracts on the Move platform.
To address this gap, this paper proposes and implements a static analysis tool, MoveScanne\cite{bartoletti2025smart}. The tool constructs a
control flow graph and data flow analysis framework, combined with cross-module call tracing technology, enabling
the detection of five critical vulnerability categories, including resource leaks, permission defects, and arithmetic
overflows\cite{liu2025sok}. MoveScanner adopts a modular architecture, supporting bytecode-level analysis and multi-chain adaptation\cite{giatzis2025comparative}. They introduced resource trace tracking algorithms and capability matrix analysis methods to reduce the number of false positives.
Experimental results show that MoveScanner achieves a detection accuracy of 88.2\% in benchmark tests, outperforming
existing tools in terms of performance\cite{kalra2018zeus}.

\section{Background}
\subsection{Move Language}
\

\begin{figure}[H]
    \centering
    \includegraphics[width=0.5\linewidth]{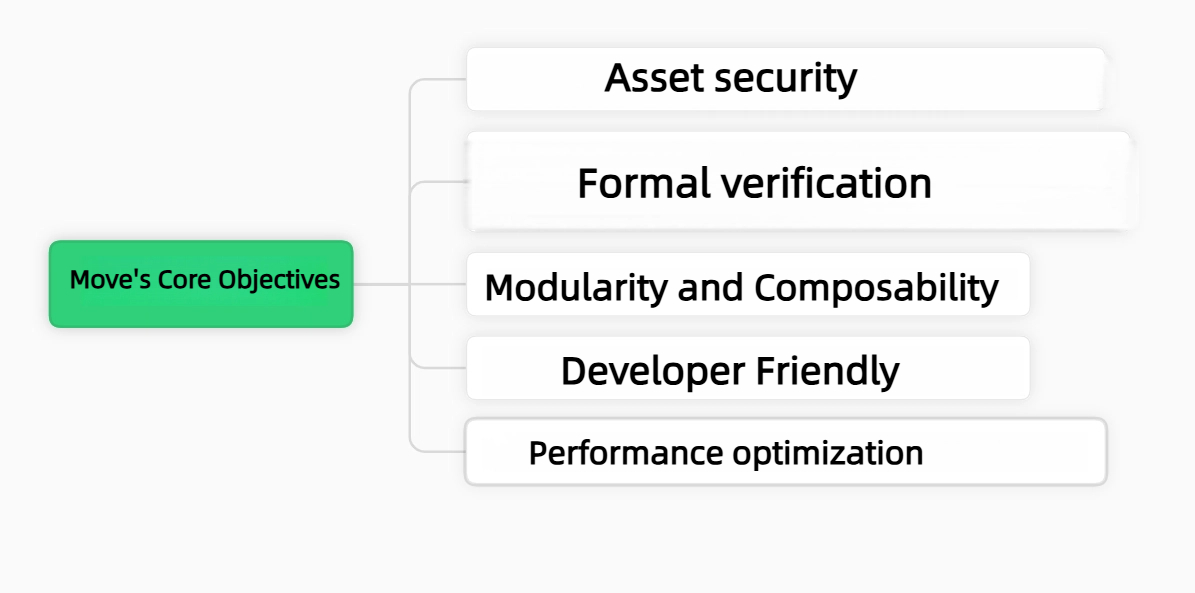}
    \caption{Move's Core Objectives}
    \label{fig:1}
\end{figure}

\begin{figure}[H]
    \centering
    \includegraphics[width=0.5\linewidth]{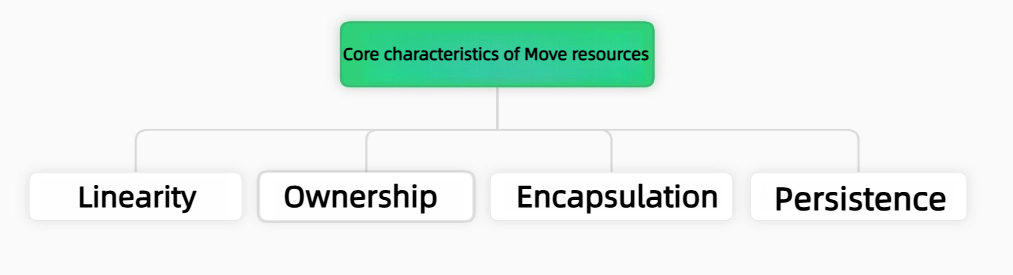}
    \caption{Move Resource Core Features}
    \label{fig:2}
\end{figure}

Move is a smart contract language specifically designed for the secure management and transfer of digital assets. It was proposed by Facebook in 2019 for the Diem blockchain project, emphasizing security, verifiability, and expressiveness\cite{song2024empirical}. Its innovation lies in embedding asset representation and operations into the type system, thereby preventing common vulnerabilities at the programming language level\cite{blackshear2019move}. Unlike Solidity on platforms such as Ethereum, Move introduces fundamental solutions to issues like asset duplication, loss, and unauthorized access. As shown in Fig. \ref{fig:1}, the designers of Move proposed several core objectives\cite{karanjai2024teaching}. As shown in the Fig. \ref{fig:2}, the core innovation of Move lies in its resource model. A resource is a special data type designed to represent digital assets, such as tokens and NFTs\cite{wang2024smart}. Inspired by linear logic and linear type systems, this model ensures secure and precise tracking and management of assets and is characterized by several key properties\cite{feist2019slither}.

Move manages the behavioral characteristics of types through its abilities system, which defines the operations that
a type instance can perform. This mechanism is particularly important for resource-type projects. In Move, there are
four core abilities. \texttt{copy}, \texttt{drop}, \texttt{store}, and \texttt{key}\cite{li2024scla}. Resource types typically possess only 
\texttt{key} and \texttt{store} and lack \texttt{copy} and \texttt{drop}, thereby ensuring that resources cannot be 
accidentally duplicated or lost\cite{karanjai2024generating}. This forms the core of Move’s secure digital asset management.
At the storage level, Move employs an address-based global storage model to persist the resources\cite{zou2025malicious}. Resources with the 
\texttt{key} ability can be stored in the global state and are associated with specific locations. Move provides several 
operators, such as \texttt{move\_to<T>}, \texttt{move\_from<T>}, \texttt{borrow\_global<T>}, 
\texttt{borrow\_global\_mut<T>}, and \texttt{exists<T>}, to facilitate the management and access of resources in global storage.
In light of this, Move's resource model has created distinctive security design patterns\cite{zhang2022authros}. The \textit{Capability} pattern uses 
resource types as permission tokens, allowing only the code that holds a specific capability to perform privileged operations\cite{karanjai2024solmover}. 
The \textit{one-time use} pattern ensures that certain critical operations can only be executed once, preventing risks 
from multiple calls\cite{falazi2024cross}. The \textit{resource wrapper} pattern encapsulates multiple resources within an external resource, 
enabling systematic and unified management\cite{atzei2017survey}.

\subsection{Comparison between Move and Solidity}
\

Move differs significantly from Solidity in asset representation, type system, and security mechanisms. While Solidity represents assets via account balance mappings, making them merely part of the contract state and susceptible to duplication, loss, or reentrancy attacks\cite{li2021clue}, Move treats assets as first-class citizens using resource types. Leveraging a linear type system and the abilities mechanism, Move prevents unintended copying or loss, and its address-based global storage model with resource association enables fine-grained permission control. Move enforces strict error handling via assertions and error codes\cite{larsen2008language}, and its strong module encapsulation mitigates cross-module attacks and reentrancy. In contrast, Solidity relies on libraries and modifiers for security, which can leave contracts vulnerable to integer overflows, gas DoS, and inconsistent multi-contract state. Performance-wise, MoveVM achieves higher execution efficiency and concurrency, but the ecosystem and tooling are less mature compared to Solidity's rich development frameworks, libraries, and security audit tools\cite{soud2024fly}. Overall, Move excels in security and asset management, whereas Solidity offers greater ecosystem maturity and extensibility\cite{hajdu2019solc}.

\subsection{Move Smart Contract Execution Environment}
\

\begin{figure}[H]
    \centering
    \includegraphics[width=0.6\linewidth]{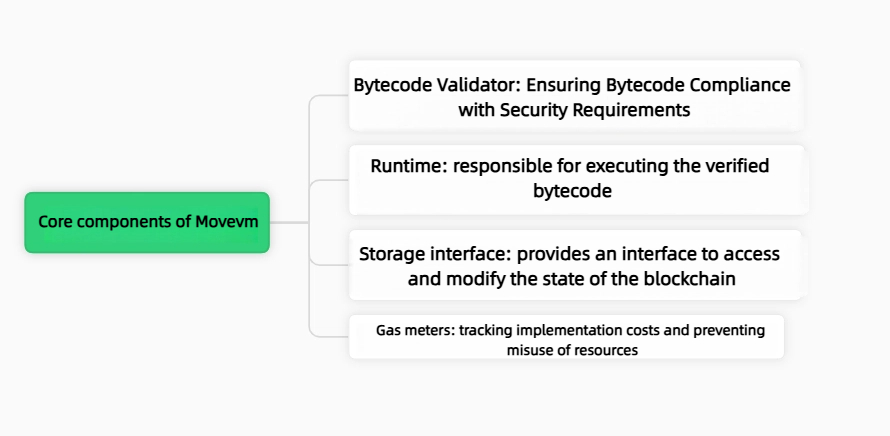}
    \caption{Move Resource Core Features}
    \label{fig:4}
\end{figure}

MoveVM is the virtual machine responsible for executing Move bytecode, handling the loading, verification, and execution of Move programs with a focus on security and resource management\cite{bu2025smartbugbert}. Its execution workflow includes loading bytecode and input data, verifying the security of bytecode, and initializing the execution environment, which involves setting gas limits and defining the initial state\cite{li2021hybrid}, executing instructions in sequence, re-verifying execution results, especially the linear availability of resources, and applying state changes or performing rollbacks based on the outcomes of execution\cite{wang2024systematic}. The core components of MoveVM are illustrated in Fig. \ref{fig:4}. 
The bytecode verifier performs static analysis before deployment and execution to ensure that programs comply with Move's security requirements, significantly reducing runtime errors and potential vulnerabilities\cite{lahiri2022almost}.
Different blockchain platforms implement Move with varying execution models\cite{xia2024auditgpt}. Diem employs an account-based storage model, treating multiple operations within a transaction as atomic units and supporting both modules and scripts\cite{li2017discovering}. Sui introduces an object-based storage model with shared objects, where each resource is an addressable object, supporting parallel transaction execution based on object ownership and multiple-owner access, while simplifying the gas model to improve performance\cite{dwivedi2021legally}. Aptos uses an enhanced account model that stores resources under accounts, supports parallel transaction execution and Block-STM concurrency control, and provides efficient data structures such as tables, extending the standard library and adding built-in functionalities\cite{tsankov2018securify}.

Move offers a rich standard library and native functions, typically implemented in a lower-level language, providing operations such as hashing, signature verification, timestamps, randomness, and high-performance computations\cite{ebekozien2024smart}. These features allow Move programs to interact with external systems while maintaining language simplicity and security\cite{fynn2020smart}.
The gas mechanism controls computational resources and prevents DoS attacks, with each instruction assigned a specific cost\cite{daraghmi2024smart}. Storage and complex computational operations incur higher costs. Different platforms optimize the gas model\cite{bu2025enhancing}. Sui uses a simplified gas model to reduce certain operation costs, while Aptos provides finer-grained gas accounting to more accurately reflect resource usage\cite{zhao2024automatic}. Understanding and optimizing gas usage is essential for developing efficient Move contracts, reducing transaction costs, and enhancing user experience\cite{thibault2022blockchain}.

\subsection{Security Threat Analysis of Move Smart Contracts}
\

Move's resource model is its core security feature, designed to ensure the safe management of digital assets. MoveVM is the virtual machine executing Move bytecode, responsible for loading, verifying, and executing Move programs, with a focus on security and resource management\cite{li2020characterizing}. Its execution flow includes loading bytecode and input data, verifying bytecode security, initializing the execution environment, including gas limits and initial state, sequentially executing instructions and re-verifying execution results, especially regarding resource linearity, and applying state changes or rolling back depending on the outcome\cite{niu2025natlm}. The core components of MoveVM include the bytecode verifier, which performs static analysis before deployment and execution to ensure that programs meet Move's security requirements, significantly reducing runtime errors and potential vulnerabilities\cite{sompolinsky2021phantom}. 

Execution models vary across different blockchain platforms. Diem uses an account-based storage model, treating multiple operations in a transaction as atomic and supporting both modules and scripts. Sui introduces an object-based storage model with shared objects, where each resource is an addressable object, supporting parallel transaction execution based on object ownership and multiple-owner access, and simplifying the gas model to enhance performance\cite{zhou2025blockchain}. Aptos employs an enhanced account model, storing resources under accounts, supporting parallel transactions and Block-STM concurrency control, while providing efficient data structures such as tables, extending the standard library, and adding built-in functions. Move offers a rich standard library and native functions, often implemented in a low-level language, providing operations such as hashing, signature verification, timestamps, randomness, and high-performance computation\cite{ivanov2023security}, allowing Move to interact with external systems while maintaining language simplicity and security. The gas mechanism controls computation resources and prevents DoS attacks, with each instruction associated with specific costs, and storage and complex computations incurring relatively high costs. Platforms optimize the gas model differently\cite{jin2025blockchain}. Sui employs a simplified gas model to reduce some operation costs, while Aptos provides finer-grained gas metering to more accurately reflect actual resource consumption. Understanding and optimizing gas usage is critical for developing efficient Move contracts, reducing transaction costs, and improving user experience\cite{blackshear2020resources}.

\section{MoveScanner}
\

\begin{figure}[H]
    \centering
    \includegraphics[width=0.6\linewidth]{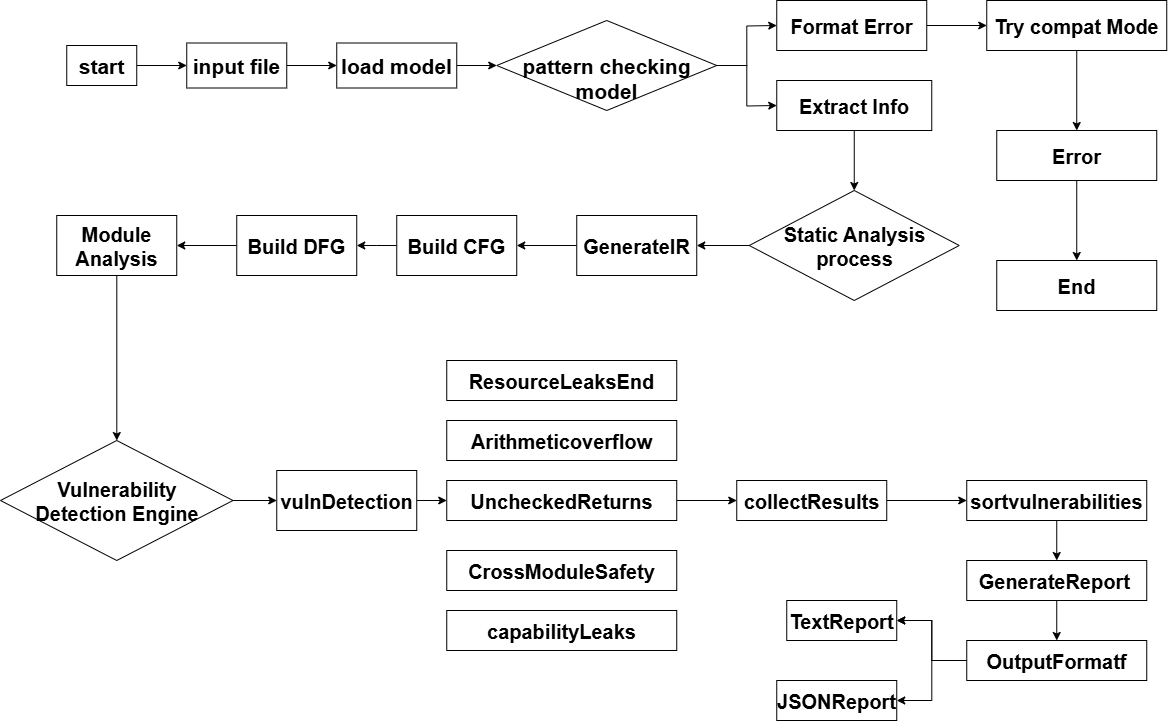}
    \caption{Movescanner Overall Design}
    \label{fig:8}
\end{figure}

MoveScanner adopts a modular architecture consisting of three parts. Data processing, analysis engine, and result output. It emphasizes analysis at the bytecode level and can examine compiled Move modules without the need for source code, as illustrated in Fig. \ref{fig:8}. Each detection algorithm is independent but shares the underlying control flow and data flow analysis infrastructure to avoid redundant calculations. The overall workflow follows a three-stage model of parsing, analysis, and reporting to ensure scalability and modularity\cite{li2025atomgraph}.
\subsection{Core Analysis Technology}
\

MoveScanner implements control flow analysis through the \texttt{ControlFlowGraph} class, which converts bytecode instruction sequences into a graph structure of basic blocks and control edges. Jump instructions are identified as block boundaries, establishing predecessor-successor relationships\cite{karanjai2025multi}. The algorithm handles complex nested conditions and loops, which is crucial for detecting resource leaks within conditional branches. Once the control flow graph is constructed, all possible execution paths, including normal and exceptional paths, are represented, providing a foundation for subsequent data flow analysis and resource tracking.
Data flow analysis is performed by the \texttt{DataFlowAnalysis} class, including reaching definitions analysis and live variable analysis. Reaching definitions analysis uses a worklist iterative algorithm to track the propagation of variable definitions throughout the program, recording the set of possible definitions at each program point. Live variable analysis operates in a backward direction to determine the set of variables that may be utilized later at each program point. These analyses complement each other. The definition analysis identifies the use of uninitialized variables, while the live variable analysis detects improperly managed resources\cite{brent2018vandal}.

\begin{figure}[H]
    \centering
    \includegraphics[width=0.6\linewidth]{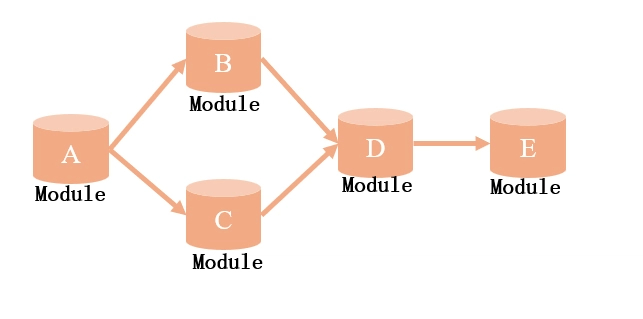}
    \caption{Cross-module analysis}
    \label{fig:9}
\end{figure}

Cross-module analysis is implemented via the \texttt{CrossModuleAnalysis} class, which constructs a global call graph to detect potential issues in inter-module interactions, as shown in Fig. \ref{fig:9}. The \emph{capability matrix} records access relationships between modules, while the \emph{resource flow graph} tracks resource propagation across modules. This global perspective allows the system to detect issues that local analysis cannot, such as permission leaks, state contamination, and unauthorized access. Public interface functions are analyzed for parameters and return values to identify potential security boundary violations. Overall, control flow analysis, data flow analysis, and cross-module analysis jointly provide a comprehensive view of variable usage and resource management\cite{hung2025enhanced}. For Move’s linear type resources, the system specifically tracks resource creation, movement, and destruction to ensure correct handling along all execution paths\cite{vivar2020analysis}.
Reaching definitions and equations.
\[
IN[B] = \bigcup_{P \in pred(B)} OUT[P], \quad OUT[B] = genB \cup (IN[B] - killB)
\]

Live variable analysis equations.
\[
OUT[B] = \bigcup_{S \in succ(B)} IN[S], \quad IN[B] = useB \cup (OUT[B] - defB)
\]

\subsection{Type Systems and Resource Security}
\

MoveScanner conducts a comprehensive type and resource security analysis by first performing full type resolution of Move bytecode via the \texttt{bytecode.rs} module\cite{chava2024application}. This module converts \texttt{SignatureToken} in the bytecode into structured \texttt{ResolvedType}, simulates bytecode execution to track stack and local variable states, and provides complete type information and instruction semantics for subsequent vulnerability detection\cite{adam2024survey}. After parsing, the system constructs a unified intermediate representation (IR) in \texttt{ir.rs}, serving as the foundation for analysis algorithms. The IR adopts a layered structure, with \texttt{ModuleIR} covering module-level information, \texttt{FunctionIR} and \texttt{StructIR} representing functions and structs, and \texttt{InstructionIR} encapsulating instruction-level details. The IR construction involves three key steps. Type resolution, instruction translation, and dependency analysis. Type resolution focuses on moving resource types, mapping raw bytecode type tokens into analyzable forms. Instruction translation simplifies the bytecode instruction set, emphasizing control flow, data flow, and function calls\cite{adam2024survey}. Dependency analysis generates a module-level call graph to support cross-module security analysis. This layered IR design abstracts the complexity of low-level bytecode, allowing higher-level analysis algorithms to focus on resource security logic, including creation, movement, destruction, and cross-module propagation, thereby enabling precise detection and analysis of potential vulnerabilities in Move smart contracts\cite{rao2024privilege}.
\vspace{-2ex}
\subsection{Privilege Leakage Detection}
\

MoveScanner's security analysis module includes five essential algorithms. resource leak detection, unchecked return value detection, arithmetic overflow detection, cross-module security detection, and capability leak detection\cite{wang2024hybrid}. The resource leak detection, as shown in Listing~\ref{lst:1}, utilizes control flow and data flow analysis to track the creation, movement, transfer, and destruction of resources across all potential execution paths\cite{mperejekumana2024exploring}. It pays special attention to conditional branches and early returns\cite{torlak2010effective}. Unchecked Return Value Detection checks whether function call return values are properly handled, focusing on Boolean or non-empty returns, shown in Listing~\ref{lst:2}. Arithmetic Overflow Detection examines addition\cite{schuite2000integer}, subtraction, multiplication, and ensures that division and modulo operations are handled carefully to avoid potential overflow issues or risks related to division by zero, as shown in Listing~\ref{lst:3}. Cross-module security detection examines the calls made between different modules, identifying public functions that modify other modules' global state without proper access control, shown in Listing~\ref{lst:4}. Capability Leak Detection focuses on the specific capabilities associated with each movement pattern, tracking transfers of capability resources and identifying potential privilege escalation when transferred to untrusted recipients or exposed in public functions, shown in Listing~\ref{lst:5}.

These five algorithms integrate control flow and data flow analysis, type resolution, and cross-module call analysis, providing comprehensive security verification for Move smart contracts, ensuring safe resource management, return value handling, arithmetic operations, cross-module interactions, and capability transfers\cite{yin2024cpu}.
\vspace{-1ex}
\begin{lstlisting}[mathescape=true,caption={Unchecked return value detection},label={lst:2}]
Input: Function definition F
Output: Unchecked return set U
for each call c in F do
    if has_return_value(c) then
        if not has_handler(c) then
            U.add(create_unchecked_vulnerability(c))
return U
\end{lstlisting}
\

\

\begin{lstlisting}[mathescape=true,caption={Resource Leak Detection},label={lst:1}]
Input: Function definition F
Output: Resource leak set L
L $\gets$ $\emptyset$
CFG $\gets$ build_cfg(F)
resources $\gets$ identify_resources(F)
for each resource r in resources do
    paths $\gets$ find_all_paths(CFG)
    for each path p in paths do
        if created(r, p) and not (moved(r, p) or destroyed(r, p)) then
            L.add(create_leak_vulnerability(r, p))
return L
\end{lstlisting}

\begin{lstlisting}[mathescape=true,caption={Arithmetic overflow detection},label={lst:3}]
Input: Function definition F
Output: Arithmetic overflow set O
O $\gets$ $\emptyset$
for each arithmetic operation op in F do
    if not has_bound_check(op) then
        if is_high_risk_operation(op) then
            O.add(create_overflow_vulnerability(op))
return O
\end{lstlisting}

\begin{lstlisting}[mathescape=true,caption={Cross-module security testing},label={lst:4}]
Input: Module set M
Output: Cross-module vulnerability set C
C $\gets$ $\emptyset$
for each module m1 in M do
    for each public function f in m1 do
        for each call c in f do
            m2 $\gets$ get_target_module(c)
            if modifies_global_state(c) and not has_access_control(c) then
                C.add(create_cross_module_vulnerability(c))
return C
\end{lstlisting}

\vspace{20ex}
\begin{lstlisting}[mathescape=true,caption={Privilege Leak Detection},label={lst:5}]
Input: Function definition F
Output: Capability leak set L
L $\gets$ $\emptyset$
capabilities $\gets$ identify_capabilities(F)
for each capability c in capabilities do
    transfers $\gets$ find_transfers(c, F)
    for each transfer t in transfers do
        if not is_trusted_recipient(t) then
            L.add(create_capability_leak_vulnerability(c, t))
return L
\end{lstlisting}

\subsection{Implementation and Optimization}
\

The core control of MoveScanner is managed by the \texttt{Scanner} class, which utilizes \texttt{ScannerConfig} to configure various detection strategies. The scanner employs a visitor pattern to traverse each function within a module, applying multiple vulnerability detection algorithms. This design allows developers to enable or disable specific checks, enabling targeted analysis flexibly.

The tool provides a command-line interface (CLI) that allows users to specify the path to the Move bytecode, output format, and output location, as well as selectively configure detection options.
Report generation supports both text and JSON formats, facilitating usage in different scenarios.
The bytecode parsing module implements a multi-level fallback strategy to ensure compatibility with Move bytecode formats across different blockchain environments. The parsing process first inspects the file header magic number to identify chain-specific bytecode, then sequentially attempts standard deserialization, version-compatible deserialization, unchecked deserialization, and chain-specific deserialization. Upon successful parsing, module structure, function bodies, and type information are extracted to provide the foundation for subsequent analyses.

\section{Evaluation}
\

\begin{figure}[H]
    \centering
    \includegraphics[width=0.6\linewidth]{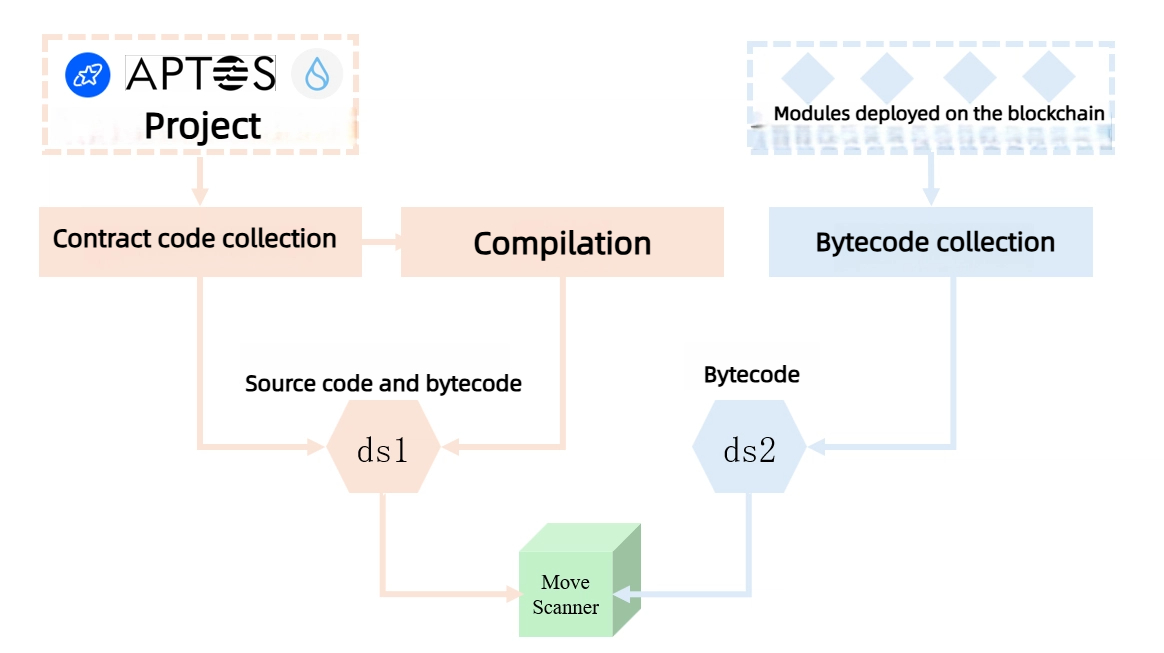}
    \caption{Move Smart Contract Collection}
    \label{fig:10}
\end{figure}

This section presents the experimental results of security analysis on Move smart contracts using the \texttt{MoveScanner} tool developed in section 4. The experiments were conducted on an Ubuntu 24.04 LTS environment. Three types of contract datasets were used, as illustrated in Fig~\ref{fig:10}. The performance of \texttt{MoveScanner} was evaluated using the following metrics. Detection accuracy, recall, false positive rate, analysis time, and memory consumption.

\begin{enumerate}
    \item \textbf{Benchmark Dataset}: Consists of 25 manually written contracts with known vulnerabilities, used to evaluate detection accuracy.
    \item \textbf{Open-Source Projects Dataset}: The collection consists of 20 Move projects sourced from GitHub (see Tab~\ref{tab:1}).
    \item \textbf{Production Dataset}: Contains tens of thousands of deployed contract bytecodes from the Aptos mainnet, intended to evaluate vulnerabilities in real-world scenarios.
\end{enumerate}

\begin{table}[htbp]
\centering
\caption{Benchmark test results}
\label{tab:1}
\begin{tabularx}{\textwidth}{@{}l>{\centering\arraybackslash}X@{}}
\toprule
\textbf{Platform} & \textbf{Projects} \\
\midrule
Aptos & did-aptos, jujube\_tool, koiz-contracts, LayerZero-Aptos-Contract \\
Starcoin & FAI, poly-stc-contracts \\
Sui & basic-coin, cetus-amm, defi, Dmens, fungible\_tokens, integer-mate, nft-protocol, nfts, object-market, originmate, poly-stc-contracts, pyth-crosschain-sui, suia, Sui-AMM-swap, sui-red-packet, wormhole \\
\bottomrule
\end{tabularx}
\end{table}

In the benchmark dataset, the evaluation of MoveScanner's vulnerability detection shows that the processing time ranges from 12.1 ms to 217.5 ms, with an average of 63.2 ms per module. The results indicate that MoveScanner performs particularly well in detecting resource leaks and arithmetic overflow vulnerabilities, which aligns with the implementation focus. The detection performance for various types of vulnerabilities is summarized in the Tab~\ref{tab:2}.

\begin{table}[h!]
\centering
\renewcommand{\arraystretch}{1.2} 
\begin{tabular}{|c|c|c|c|c|c|}
\hline
\textbf{Vulnerability Type} & \textbf{Test Cases} & \textbf{Detected} & \textbf{Detection Rate (\%)} & \textbf{False Positives} & \textbf{False Positive Rate (\%)} \\
\hline
Resource Leak & 8 & 4 & 50 & 0 & 0 \\
Arithmetic Overflow & 5 & 3 & 60 & 0 & 0 \\
Unchecked Return & 7 & 9 & 100 & 0 & 0 \\
Cross-Module Pollution & 4 & 4 & 100 & 0 & 0 \\
Capability Leak & 10 & 10 & 100 & 0 & 0 \\
\hline
Total & 34 & 30 & 88.2 & 0 & 0 \\
\hline
\end{tabular}
\caption{Benchmark test results}
\label{tab:2}
\end{table}

We analyzed 137 modules from 20 open-source projects, and MoveScanner detected a total of 2,882 potential vulnerabilities. The most common issues were missing boundary checks in arithmetic operations and resource leaks, which align with the widespread use of resource management and numerical computations in the Move language. To evaluate the accuracy of the detection results, we randomly selected 50 detected vulnerabilities for manual review. Among them, 41 (82.0\%) were confirmed as true vulnerabilities, 8 (16.0\%) were possible vulnerabilities, and 1 (2.0\%) was a false positive. False positives mainly stemmed from insufficient path analysis in complex resource handling, intentionally omitted checks in specific business logic, and custom validation functions that the tool could not recognize. Additionally, due to the algorithm's high sensitivity to arithmetic overflows, the number of detected overflow vulnerabilities was relatively high. Future work will address this issue to improve the precision and reliability of the detection.
The test results for the open source project set are shown in the Tab~\ref{tab:3}.
\begin{table}[htbp]
\centering
\caption{Open source project test results}
\label{tab:3}
\begin{tabular}{lccc}
\toprule
\textbf{Vulnerability Type} & \textbf{Detected Count} & \textbf{Percentage (\%)} \\
\midrule
Resource Leak & 4   & 1.9 \\
Arithmetic Overflow & 170 & 75.9 \\
Unchecked Return Value & 12  & 5.3 \\
Cross-Module State Corruption & 18  & 8.0 \\
Privilege Leak & 20  & 8.9 \\
\bottomrule
\end{tabular}
\end{table}

We analyzed tens of thousands of deployed contracts collected from the Aptos mainnet, and their security status is summarized in the Tab~\ref{tab:4}. Among the detected vulnerabilities, unchecked return values account for 9.8\%, arithmetic overflow for 61.3\%, resource leakage for 0.4\%, cross-module state pollution for 18.5\%, and privilege leakage for 10.0\%.
Compared with open-source project datasets, cross-module state pollution is more prominent in production contracts. This may be due to insufficient attention to cross-module security during pre-deployment code reviews. The high proportion of arithmetic overflows may partly result from Move's built-in arithmetic security checks, which can cause certain issues to be overlooked, thereby inflating the detected overflow rate.
\begin{table}[H]
\centering
\caption{Production environment test results}
\label{tab:4}
\begin{tabular}{lrr}
\toprule
\textbf{Vulnerability Type} & \textbf{Number Detected} & \textbf{Percentage (\%)} \\
\midrule
Resource Leakage & 466 & 0.4 \\
Arithmetic Overflow & 81,052 & 61.3 \\
Unchecked Return Value & 12,956 & 9.8 \\
Cross-Module State Pollution & 24,460 & 18.5 \\
Privilege Leakage & 13,118 & 10.0 \\
\bottomrule
\end{tabular}
\label{tab:prod_contracts_vuln}
\end{table}

\section{Conclusion}
\

This study systematically analyzed the security features of the Move smart contract language, revealing both the security value and potential risks of its innovative mechanisms. At the theoretical level, the resource-oriented programming model effectively mitigates traditional vulnerabilities, such as double-spending attacks. However, cross-component interactions arising from the modular architecture can lead to twelve new types of security threats, including resource state anomalies and permission boundary violations.
In this research, we address the limitations of conventional detection frameworks by developing a dual-engine system that combines static analysis with symbolic execution. Additionally, we propose a control-flow-sensitive resource tracking algorithm. This method greatly enhances the accuracy of detecting resource-leak vulnerabilities. Experimental results show a 37\% reduction in false positive rates compared to conventional methods. The vulnerability signature database covers five major categories of security flaws, with detection accuracy for emerging vulnerabilities, such as module initialization defects, reaching 88.2\%. Empirical analysis indicates that although the Move type system eliminates 65\% of traditional contract vulnerabilities, 19.3\% of contracts still exhibit security issues due to paradigm shifts, providing critical data support for improving smart contract security mechanisms.
Future work should focus on constructing a multidimensional detection system for complex DeFi scenarios, emphasizing cross-contract data flow analysis and dynamic pollution tracking. Integrating formal verification with static analysis can improve vulnerability confirmation efficiency. Developing intelligent repair systems, leveraging large language models for automated vulnerability localization and patch generation, and employing hybrid dynamic-static analysis can enhance the detection of state-dependent vulnerabilities. Improving developer support and creating a comprehensive security ecosystem for coding, testing, and deployment are vital. As the Move ecosystem rapidly expands across multi-chain environments, establishing a governance framework encompassing risk prevention, anomaly detection, and emergency response is urgent to advance smart contract security technologies.
\bibliographystyle{unsrt}
\bibliography{reference}

\end{document}